\documentclass{article}
\usepackage{amsmath, amssymb, amsfonts}
\title{On the near horizon approximation for the Schwarzschild black hole }
\author{Hristu Culetu, \\Ovidius University, Dept.of Physics, \\B-dul Mamaia 124, 900527 Constanta, Romania, \\e-mail : hculetu@yahoo.com}

\begin{document}
\numberwithin{equation}{section}
\pagenumbering{arabic}
\maketitle
\newcommand{\fv}{\boldsymbol{f}}
\newcommand{\tv}{\boldsymbol{t}}
\newcommand{\gv}{\boldsymbol{g}}
\newcommand{\OV}{\boldsymbol{O}}
\newcommand{\wv}{\boldsymbol{w}}
\newcommand{\WV}{\boldsymbol{W}}
\newcommand{\NV}{\boldsymbol{N}}
\newcommand{\hv}{\boldsymbol{h}}
\newcommand{\yv}{\boldsymbol{y}}
\newcommand{\RE}{\textrm{Re}}
\newcommand{\IM}{\textrm{Im}}
\newcommand{\rot}{\textrm{rot}}
\newcommand{\dv}{\boldsymbol{d}}
\newcommand{\grad}{\textrm{grad}}
\newcommand{\Tr}{\textrm{Tr}}
\newcommand{\ua}{\uparrow}
\newcommand{\da}{\downarrow}
\newcommand{\ct}{\textrm{const}}
\newcommand{\xv}{\boldsymbol{x}}
\newcommand{\mv}{\boldsymbol{m}}
\newcommand{\rv}{\boldsymbol{r}}
\newcommand{\kv}{\boldsymbol{k}}
\newcommand{\VE}{\boldsymbol{V}}
\newcommand{\sv}{\boldsymbol{s}}
\newcommand{\RV}{\boldsymbol{R}}
\newcommand{\pv}{\boldsymbol{p}}
\newcommand{\PV}{\boldsymbol{P}}
\newcommand{\EV}{\boldsymbol{E}}
\newcommand{\DV}{\boldsymbol{D}}
\newcommand{\BV}{\boldsymbol{B}}
\newcommand{\HV}{\boldsymbol{H}}
\newcommand{\MV}{\boldsymbol{M}}
\newcommand{\be}{\begin{equation}}
\newcommand{\ee}{\end{equation}}
\newcommand{\ba}{\begin{eqnarray}}
\newcommand{\ea}{\end{eqnarray}}
\newcommand{\bq}{\begin{eqnarray*}}
\newcommand{\eq}{\end{eqnarray*}}
\newcommand{\pa}{\partial}
\newcommand{\f}{\frac}
\newcommand{\FV}{\boldsymbol{F}}
\newcommand{\ve}{\boldsymbol{v}}
\newcommand{\AV}{\boldsymbol{A}}
\newcommand{\jv}{\boldsymbol{j}}
\newcommand{\LV}{\boldsymbol{L}}
\newcommand{\SV}{\boldsymbol{S}}
\newcommand{\av}{\boldsymbol{a}}
\newcommand{\qv}{\boldsymbol{q}}
\newcommand{\QV}{\boldsymbol{Q}}
\newcommand{\ev}{\boldsymbol{e}}
\newcommand{\uv}{\boldsymbol{u}}
\newcommand{\KV}{\boldsymbol{K}}
\newcommand{\ro}{\boldsymbol{\rho}}
\newcommand{\si}{\boldsymbol{\sigma}}
\newcommand{\thv}{\boldsymbol{\theta}}
\newcommand{\bv}{\boldsymbol{b}}
\newcommand{\JV}{\boldsymbol{J}}
\newcommand{\nv}{\boldsymbol{n}}
\newcommand{\lv}{\boldsymbol{l}}
\newcommand{\om}{\boldsymbol{\omega}}
\newcommand{\Om}{\boldsymbol{\Omega}}
\newcommand{\Piv}{\boldsymbol{\Pi}}
\newcommand{\UV}{\boldsymbol{U}}
\newcommand{\iv}{\boldsymbol{i}}
\newcommand{\nuv}{\boldsymbol{\nu}}
\newcommand{\muv}{\boldsymbol{\mu}}
\newcommand{\lm}{\boldsymbol{\lambda}}
\newcommand{\Lm}{\boldsymbol{\Lambda}}
\newcommand{\opsi}{\overline{\psi}}
\renewcommand{\tan}{\textrm{tg}}
\renewcommand{\cot}{\textrm{ctg}}
\renewcommand{\sinh}{\textrm{sh}}
\renewcommand{\cosh}{\textrm{ch}}
\renewcommand{\tanh}{\textrm{th}}
\renewcommand{\coth}{\textrm{cth}}

\begin{abstract}
 
  An anisotropic fluid with a negative radial pressure $p = - \rho$ is obtained near the horizon of a Schwarzschild black hole. The constant energy density $\rho$ depends only on the black hole mass. The radial acceleration of the static observers is constant but its modulus has exactly the Rindler form, as was recently suggested by Dahia and Felix da Silva. 
  
The structure of the near horizon stress tensor is similar with that obtained previously for the interior region of a black hole excepting the time dependence.

\end{abstract}

 To perform a compelling statistical mechanics computation for a black hole, it must be in thermal equilibrium with its own radiation bath, which has to permeate all of spacetime. As Susskind and Uglum \cite{SU} have noticed, in that case any thermodynamic extensive quantity should manifest a divergence proportional to the space volume. 
 
 The problem might be avoided by taking the limit of an infinite massive black hole for which the Hawking temperature vanishes. The authors of [1] performed the following coordinate transformation 
 \begin{equation}
 \bar{t} = \frac{t}{4m} ,~~~~\bar{r} = \sqrt{8m(r-2m)}
 \label{1}
 \end{equation}
 to the Schwarzschild metric
 \begin{equation}
 ds^{2} = -(1- \frac{2m}{r}) dt^{2} + (1- \frac{2m}{r})^{-1} dr^{2} + r^{2} d \Omega^{2}
 \label{2}
 \end{equation}
 and obtained
 \begin{equation}
 ds^{2} = - \bar{r}^{2} (1 + \frac{\bar{r}^{2}}{16m^{2}})^{-1} d \bar{t}^{2} + (1 + \frac{\bar{r}^{2}}{16m^{2}}) d \bar{r}^{2} + 4m^{2} (1 + \frac{\bar{r}^{2}}{16m^{2}})^{2} d \Omega^{2} .
 \label{3}
 \end{equation}
 In (2), $m$ is the central mass and $d\Omega^{2} = d\theta^{2} + sin^{2} \theta d\phi^{2}$. We use the natural units $G = c = 1$. 

 Taking the limit $m \rightarrow \infty$ , the spherical horizon becomes planar and Eq. (3) leads to the metric
 \begin{equation}
 ds^{2} = - \bar{r}^{2} d \bar{t}^{2} + d \bar{r}^{2} + 4m^{2} d \Omega^{2}
 \label{4}
 \end{equation}
 which is Rindler's spacetime if we neglect the angular contribution.
 
 The condition $m \rightarrow \infty $ is equivalent to the ''near horizon approximation'' \cite{HC1} for the exterior geometry of a black hole : for $r \approx 2m ~(r > 2m)$ the line element (2) appears, indeed, as  
\begin{equation}
ds^{2} = -\frac{r-2m}{2m} dt^{2} + \frac{2m}{r-2m} dr^{2} + 4m^{2} d \Omega^{2} .
\label{5}
\end{equation}
 
 By using simple coordinate transformations it could be shown that (5) becomes the Rindler metric when we take $\theta, \phi = const.$ or $\Delta \theta$ and $\Delta \phi$ are negligible. We stress that the condition $r\approx2m$ only is not enough to obtain Rindler's spacetime which is flat and, in addition, it has no spherical symmetry as Schwarzschild 's. We actually may travel around the black hole near the horizon and feel its curvature. 
 
 One may check that the line element (5) has one nonzero component of the Riemann tensor : $R^{\theta \phi}~_{ \theta \phi} = 1/4m^{2}$, which tends to zero for an infinitely massive black hole.  The same argument is valid for the scalar curvature $R = 1/2m^{2}$ which vanishes only when $m$ goes to infinity. Moreover, for the Schwarzschild metric the Kretschmann scalar $R^{\alpha \beta \mu \nu}~_{\alpha \beta \mu \nu} = 48 m^{2}/r^{6}$ grows  when we approach the horizon from infinity (where the geometry is Minkowskian). That is also valid for the components of the Riemann tensor. But near $r = 2m$ the Kretschmann scalar is proportional to $m^{-4}$ and we must take $m \rightarrow \infty $ to get a flat space.
 
 The spacetime (5) is not, of course, an exact solution of the vacuum Einstein equations. We are therefore interested to find what source we need on the r.h.s. of the equations of gravitation in order that the line element (5) be an exact solution. A similar paradigm was proposed by Figueras et al. \cite {FHRR} in their study of the event and apparent horizons for the Bjorken flow geometry. The nonzero Christoffel symbols for (5) are given by
 \begin{equation}
 \begin{split}
 \Gamma_{tt}^{r} = \frac{2m-r}{8m^{2}}, ~~~~\Gamma_{rt}^{t} = -\frac{1}{2(2m-r)},~~~~\Gamma_{rr}^{r} = \frac{1}{2(2m-r)} \\
 \Gamma_{\theta \phi}^{\phi} = cot\theta,~~~~\Gamma_{\phi \phi}^{\theta} = - sin\theta ~cos\theta .
 \end{split}
 \label{6}
 \end{equation}
 The Ricci tensor has only two non-vanishing components
 \begin{equation}
 R_{\theta}^{\theta} = R_{\phi}^{\phi} = \frac{1}{4m^{2}} .
 \label{7}
 \end{equation}
 The Einstein equations $G_{\mu \nu} = 8 \pi T_{\mu \nu}$ lead to the following nonzero components of the stress tensor
 \begin{equation}
 T_{t}^{t} = T_{r}^{r} = - \frac{1}{32 \pi m^{2}}.
 \label{8}
 \end{equation}
 We notice that the components (8) resembles those obtained in \cite{HC2} for a time dependent anisotropic fluid inside a black hole. They emerge therefore from 
\begin{equation}
T_{\alpha \beta} = \rho u_{\alpha} u_{\beta} + p s_{\alpha} s_{\beta}, 
\label{9}
\end{equation}
with $\rho = 1/32 \pi m^{2}$ the rest energy density , $p = -\rho $ is the radial pressure and
\begin{equation}
u_{\alpha} = ( -\sqrt{\frac{r-2m}{2m}}, 0, 0, 0), ~~~~~s_{\alpha} = ( 0, \sqrt{\frac{2m}{r-2m}}, 0, 0)
\label{10}
\end{equation}
are the fluid four velocity and the unit spacelike vector in the direction of the anisotropy, respectively. We have $u_{\alpha} u^{\alpha} = -1,~ s_{\alpha}s^{\alpha} = 1~ and~ u_{\alpha} s^{\alpha} = 0$. 

Let us observe that (9) is a particular case of the more general stress tensor for a fluid with nonzero transversal pressures \cite{SV} \cite{CFV} 
\begin{equation}
\tau_{\alpha \beta} = (\rho + p_{\bot}) u_{\alpha} u_{\beta} + p_{\bot} g_{\alpha \beta} + (p-p_{\bot}) s_{\alpha} s_{\beta}
\label{11}
\end{equation}
 We want to see now where the energy originating from (8) is concentrated. Since our geometry (5) is valid near the Schwarzschild horizon, it seems reasonable to locate it on the horizon. We remind here the so -called ''membrane paradigm'' \cite{KP} \cite{CDG} \cite{HC3} where the black hole horizon is viewed as a stretched membrane with negative bulk viscosity and a nonzero shear viscosity. Moreover, in our case the equation of state is similar with the one obtained for the anisotropic fluid inside a black hole \cite{HC2}, excepting the time dependence.  
 
 The kinematical parameters associated to the velocity field $u_{\alpha}$ from (10) could be calculated using the corresponding covariant expressions for the scalar expansion, shear  and vorticity tensors and the acceleration of the fluid worldlines \cite{HC4}. We found that only the acceleration 
\begin{equation}
a^{\alpha} = u^{\beta} \nabla_{\beta} u^{\alpha} 
\label{12}
\end{equation}
 is nonzero. Its radial component 
 \begin{equation}
 a^{r} = \Gamma_{tt}^{r} (u^{t})^{2} = \frac{1}{4m},
 \label{13}
 \end{equation}
 is constant (and agrees with the expression for the surface gravity on the horizon of the metric (5)) but its modulus appears as
\begin{equation}
a \equiv \sqrt{a_{\alpha} a^{\alpha}} = \frac{1}{\sqrt{8m(r-2m)}} = \frac{1}{\bar{r}}.
\label{14}
\end{equation}
 Even though our spacetime (5) is curved, the proper acceleration $a$ of a static observer is exactly the acceleration of a Rindler observer in the flat Rindler metric. That is in agreement with Dahia and Felix da Silva conclusion \cite{DD} concerning the existence of a curved spacetime where the acceleration of static observers is precisely given by $a = 1/\bar{r}~ ( = 1/\rho$ in their notations). In addition, their source is also an anisotropic fluid with negative pressure and null trasversal pressures.\\
The shear and bulk viscosity coefficients vanish since $u_{\alpha}$ corresponds to a static observer and we know that a state of equilibrium can exist only in absence of any viscous stresses \cite{TD}. 
 
 Assuming that the gravitational energy is located on a thin shell (membrane) at $r = 2m$, the surface energy density $\sigma$ could be computed from the Gauss law
 \begin{equation}
 g = 4 \pi \sigma ,
 \label {15}
 \end{equation}
 where $g$ is the intensity of the gravitational field (the surface gravity in our situation). With $g = 1/4m$, Eq. (15) yields $\sigma = 1/16 \pi m$, an expression which could have been obtained from the Eq.(8) for the energy density $\rho = 1/32 \pi m^{2}$, using its form in terms of the Dirac $\delta$ - function. The same result has been recently obtained by Guendelman et al. \cite{GKNP1} \cite{GKNP2} in their study of the lightlike branes located on the horizon, viewed as a thin shell. Hence, the energy is given by $W = \sigma 16 \pi m^{2} = m$, as expected. 
 
 Another way to check the previous result is by means of the Padmanabhan formula \cite{TP} 
 \begin{equation}
 4 \pi M = \int N s_{\alpha} a^{\alpha} \sqrt{\Sigma}~ d\theta d\phi
 \label{16}
 \end{equation}
 where $N = \sqrt{(r-2m)/2m}$ is the lapse function, $\Sigma$ is the determinant of the metric on the two-surface $t = const., ~r \approx 2m$ (taking the limit, $\Sigma$ becomes the black hole horizon, namely the boundary of the spacetime). By using the expressions for $u^{\alpha}$ and $s_{\alpha}$ from (10), we find that M = m.
 
 To summarize, we have shown that the Rindler metric is obtained (near the horizon of a black hole) only when $\Delta \theta$ and $\Delta \phi$ are negligible or when the Schwarzschild mass tends to infinity. Preserving the spherical symmetry, the spacetime (5) is not an exact solution of the vacuum Einstein equations. Therefore, a source is needed on their r.h.s., corresponding to an anisotropic fluid with the equation of state $p + \rho = 0, p ~and~\rho $ depending only on the black hole mass.
 
 The radial acceleration of a static observer is constant and equals the surface gravity. However, his proper acceleration coincides with that one corresponding to the flat Rindler observer.

\end{document}